# Interplay of broken symmetry and delocalized excitations in the insulating state of 1$T$-TaS$_2$

Xun Jia[1,*], Anubhab Haldar[2], Jungho Kim[3], Yilin Wang[4], Gilberto Fabbris[3], Karl Ludwig[5,6], Stefanos Kourtis[7], Mary Upton[3], Yu Liu[8], Wenjian Lu[8], Xuan Luo[8], Yu-Ping Sun[8,9,10], Diego Casa[3], Sahar Sharifzadeh[2,5,6], Pierre T. Darancet[11,*], Yue Cao[1,12,*]

[1] *Materials Science Division, Argonne National Laboratory, Lemont, Illinois 60439, USA*
[2] *Department of Electrical and Computer Engineering, Boston University, Boston, MA, USA*
[3] *Advanced Photon Source, Argonne National Laboratory, Lemont, Illinois 60439, USA*
[4] *Hefei National Laboratory for Physical Science at Microscale, University of Science and Technology of China, Hefei 230026, China*
[5] *Department of Physics, Boston University, Boston, MA 02215, USA*
[6] *Division of Materials Science and Engineering, Boston University, Boston, MA, USA*
[7] *Institut quantique & Departement de physique, Universite de Sherbrooke, Quebec J1K 2R1, Canada*
[8] *Key Laboratory of Materials Physics, Institute of Solid State Physics, HFIPS, Chinese Academy of Sciences, Hefei 230031, China*
[9] *High Magnetic Laboratory, HFIPS, Chinese Academy of Sciences, Hefei 230031, China*
[10] *Collaborative Innovation Centre of Advanced Microstructures, Nanjing University, Nanjing 210093, China*
[11] *Center for Nanoscale Materials, Argonne National Laboratory, Lemont, Illinois 60439, USA*
[12] *Pritzker School of Molecular Engineering, University of Chicago, Chicago, Illinois 60637, USA*

Emails: xjia@anl.gov (X.J.), ssharifz@bu.edu (S. S.), pdarancet@anl.gov (P.T.D.), yue.cao@anl.gov (Y.C.)

*Abstract*

*Coexistence of localized and extended excitations is central to the macroscopic properties of correlated materials. For 5d transition metal compounds, electron correlations alone generally do not lead to a metal-insulator (Mott) transition, with insulating behavior usually resulting from their coupling with magnetic ordering and/or structural distortions. 1T-TaS$_2$ is a prototypical example of such correlated insulating behavior, with a high-symmetry metallic phase transforming into a distorted, charge density wave (CDW) insulating state at low temperatures. The relevance of the localized electron physics at play in 3d compounds to these 5d transition metal compounds remains an open question. We resolved this standing controversy in 1T-TaS$_2$ combining resonant inelastic X-ray spectroscopy and first-principles calculations. We observed five electronic excitations arising from the interband transitions of the Ta 5d orbitals and the S 3p ligand state, with none of the excitations on the order of the Mott gap. These excitations cannot be explained within the framework of standard multiplet calculations that assume a localized wavefunction, but instead, are captured by a band theory framework accounting for the low symmetry of the crystal field in the CDW state. Our findings suggest that the electronic property of 1T-TaS$_2$ is dominated by both plasmonic quasiparticles and inter-band transitions associated with a Drude-type response, with no resonance associated with a putative Mott transition. Our discovery provides new insights into the electron localization and the onset of insulating behavior in 5d transition metal materials.*

# I. Introduction

Metal-Insulator "Mott" transitions [1] canonically involve localized electronic states, such as electrons in $d$ and $f$ orbitals with low principal quantum numbers, leading to repulsive Coulomb interactions outcompeting the delocalization from the orbital overlap. Mott insulators are ubiquitous in $3d$ transition metal oxides, with the most notable example being the cuprates at the center of high $T_C$ superconductor research for the last few decades [2]. In contrast, when the nature of the underlying electronic states becomes more delocalized, e.g., in most $5d$ transition metal compounds, electronic correlations alone generally cannot lead to localization; and metal-insulator transitions usually involve additional energy scales, such as polaronic effects and spin-orbit coupling. These additional energy scales naturally raise the question of the transferability of the phenomenological models developed to describe $3d$ transition metal compounds to their heavier counterparts. In recent years, numerous studies dedicated to the iridate compounds [3, 4, 5] have shown that the combination of spin-orbit coupling and Coulomb repulsion leads to a Mott insulating behavior, most notably in $Sr_2IrO_4$ and $Sr_3Ir_2O_7$. Despite the role of spin-orbit coupling, theoretical and experimental studies have identified similarities in the behavior of iridates and cuprates both in pristine and doped materials, including electronic structures and magnetic excitations [3, 4, 6]. In contrast, the nature of the excitations in $5d$ insulators involving charge density waves (CDW) is less understood and raises the question on the transferability of the models derived from intrinsically localized electronic states.

Tantalum disulfide ($1T$-$TaS_2$) is a layered $5d$ transition compound among the larger family of transition metal dichalcogenides (TMDCs) with charge orders that are actively studied due to their proximity to superconductivity and potential applications in quasi-two-dimensional electronics [6, 8]. In Fig. 1(a), we show the crystal structure of $1T$-$TaS_2$ in the high symmetry metallic phase. The crystal is centrosymmetric with the Ta atom sitting at the center of a tilted sulfur (S) octahedron. $1T$-$TaS_2$ has a commensurate, 'star-of-David'-type CDW $Q_{CCDW} = (3/13, 1/13, 1/3)$ below 225K, a nearly commensurate CDW at $Q_{NCDW} = (0.245, 0.068, 1/3)$ below 355K and an incommensurate CDW of $Q_{ICDW} = (0.283, 0, 1/3)$ above 355K [8, 9]. The symmetry-broken phases are significantly less conductive than their high symmetry counterpart, and the transition from the NCDW into the CCDW phase features a 10-fold increase in resistivity [8].

Despite decades of research, the nature of the insulating state in the CCDW phase $1T$-$TaS_2$ is actively debated. On the one hand, the 13 Ta atoms per CCDW supercell have a nominal $5d^1$ electronic configuration [8, 10, 11], prompting the widely-accepted argument that the Coulomb repulsion plays a vital role and that the onset of insulator behavior is of Mott type. On the other hand, the band dispersion in the CCDW phase observed using the angle-resolved photoemission spectroscopy (ARPES) can be captured reasonably well through band folding without significant electron correlation. In this latter scheme, the band gap can be induced e.g., through other mechanisms such as dimerization, suggesting an in-plane band insulator with a one-dimensional out-of-plane metallic order subsequently broken by a Peierls-type or Anderson-like localization [12]. These competing pictures have been actively debated over the last few years [12, 13, 14, 15] with their fundamental difference in whether the electron wavefunction is localized or extended. Recently, $1T$-$TaS_2$ was proposed as a possible quantum spin liquid (QSL) [16] partially supported by specific heat and muon spin resonance studies [17, 18]. The QSL proposal was closely linked to the Mott picture, further highlighting the need to understand the nature of the electron wavefunctions in $1T$-$TaS_2$.

In this paper, we study the electronic excitations in $1T$-$TaS_2$ combining resonant inelastic X-ray scattering (RIXS) and theoretical simulations. For $3d$ transition metal compounds, excitations from local electronic interactions can be measured with RIXS and calculated using orbital multiplets [19]. The qualitative and quantitative agreements between experiments and orbital multiplet calculations have been well established [20, 21, 22, 23] due to the spatial localization of $3d$ orbitals. The same approach was applied to the interpretation of $5d$ transition metal compounds, notably iridates, despite the $5d$ orbitals being substantially more extended than their $3d$ counterparts [24, 25]. Our key finding is that the RIXS spectra from 1$T$-TaS$_2$ qualitatively deviate from the orbital-multiplet predictions and hence from a localized electron wavefunction theory similar to the *3d* compounds. In contrast, the RIXS excitations can be explained by orbital excitations from an extended electron wavefunction and, by taking into account the broken symmetry of the crystal field, indicate that polaronic effects dominate the response function of the insulating state at low energies. Our results suggest that the nature of 1$T$-TaS$_2$ is not a *3d*-like Mott insulator but one close to a band insulator. Moreover, our calculations demonstrate that RIXS and band-theory-based calculations can work in concert describing orbital excitations in materials with itinerant electrons.

## II.    Results

The RIXS measurements were performed at the 27-ID-B (the MERIX endstation) of the Advanced Photon Source with a total energy resolution around 63 meV. The incident X-ray was monochromatized at the Ta L$_3$ edge. A schematic of the RIXS process at the Ta L$_3$ edge is displayed in Fig. 1(b). The incident X-ray photon excites an electron from the Ta *2p* core level to the empty *5d* levels above the Fermi level. Electrons in the Ta *5d* levels subsequently recombine with the *2p* core hole and emit an X-ray photon. The energy loss between the incident and the scattered X-rays thus contains information about the low-energy electronic excitations in the material. The X-ray absorption spectra from 1$T$-TaS$_2$ and the reference Ta metal are plotted in Fig. 1(c). In Fig. 1(d) we show the RIXS spectra as a function of the incident energy (marked with vertical black lines in Fig. 1(c)) across the Ta L$_3$ edge at a base temperature between 10~15 K. Further details about the X-ray measurements as well as the material synthesis and characterization are presented in the **Supplementary Materials** [26]. To zeroth order, we observe a series of energy loss features (marked with the red arrows) within the first 10 eV, which do not depend on the incident X-ray energy. In comparison, the Ta fluorescence onsets above the Ta L$_3$ edge, corresponding to the energy loss features that increase linearly with the incident X-ray energy with a slope of unity. It is apparent that the Ta fluorescence only accounts for a very small fraction of the RIXS spectral weight. This is consistent with other Ta-containing materials e.g., GaTa$_4$Se$_8$ [26], and in contrast to the scenario in itinerant materials where the fluorescence is supposed to dominate the spectral intensity. Notably, the RIXS peaks from GaTa$_4$Se$_8$ were interpreted satisfactorily using the orbital multiplets [26]. Indeed, at the first glance, our RIXS measurements seem to suggest that 1$T$-TaS$_2$ is better described in terms of electrons with strong local correlations as expected in a typical Mott insulator.

We performed a multiple peak fitting to quantify the energy loss features in 1$T$-TaS$_2$ and compared them with the multiplet theory predictions. Fig. 2(a) displays the RIXS spectra at the incident energy of 9882 eV. This is at the peak of the Ta L$_3$ edge and where the X-ray Raman peaks are most prominent. As a comparison, in iridates, the RIXS features are the strongest at the Ir L$_3$ pre-edge, 3 eV lower than the absorption peak [6, 24, 25]. In 1$T$-TaS$_2$, five energy loss features are needed to describe the RIXS spectra. We list the position and the full width at half maximum (FWHM) of these peaks in Table I. Each energy loss peak is described using a Lorentzian while the elastic line takes the shape of a Gaussian. The resulting peak positions and FWHMs do not depend sensitively on the choice of line shapes. Most prominently, there are no clear energy loss features on the order of the reported "Mott" gap of 450 meV [15]. This lack of a Mott gap feature was further confirmed in our RIXS data collected using a Montel mirror based spectrometer with 12 meV energy resolution (see **Supplementary Materials** for details) [26], and was in sharp contrast to previous observations in established Mott insulators with localized electron wavefunctions [6, 19]. Note that the X-ray penetration depth at the Ta *L*-edge is ~ 3 μm. The RIXS signal is therefore not sensitive to the details of the surface termination or stacking. Thus, the absence of features corresponding to the Mott gap points strongly away from a Mott picture of localized electron wavefunctions.

From Table I, the four peaks with energy losses less than 6 eV have FWHMs broader than the sub-2 eV peaks but comparable to the peak around 3.5 eV in GaTa$_4$Se$_8$ [26]. The peak at 7.11 eV is significantly broader, with a FWHM over 4 eV. All these peaks are substantially broader than those in typical iridates [6, 24, 25]. This, combined with the lack of a peak at the energy of the Mott gap, indicates that the electrons in

$1T$-TaS$_2$ are substantially less localized, if at all, than those in established Mott insulators.

**Table I: Fitted peak positions and full width at half maximums (FWHMs) from the RIXS spectra.**

| Peak label | Energy (eV) | FWHM (eV) |
|---|---|---|
| 1 | 1.30 ± 0.11 | 1.02 ± 0.33 |
| 2 | 2.02 ± 0.06 | 0.91 ± 0.33 |
| 3 | 4.14 ± 0.02 | 1.18 ± 0.11 |
| 4 | 5.04 ± 0.08 | 2.10 ± 0.30 |
| 5 | 7.11 ± 0.05 | 4.21 ± 0.16 |

Indeed, a more detailed look at the RIXS spectra casts doubts over the Mott picture. A multiplet calculation was performed for a single Ta atom by constructing Wannier orbitals centered on the Ta atom. In the Mott limit, the local electronic correlation dominates with a nominal Ta 5d$^1$ configuration. As such, we argue the choice of a single Ta atom would suffice to capture the localized d-d excitations if the system is really in the Mott limit. The details of the multiplet calculation are in the **Supplementary Materials** [26]. Within the first few eVs, the energy loss features mainly arise from the crystal field splitting between the $t_{2g}$ and $e_g$ orbitals. Taking into account the sizable spin-orbit coupling (SOC) of the Ta atom, the $t_{2g}$ level further splits into J$_{eff}$ = 1/2 and J$_{eff}$ = 3/2 states with an energy difference a fraction of an eV, similar to those observed in iridates [6, 24, 25].

We show the calculated multiplet peaks without (green) and with (black) an SOC of 0.2 eV using vertical solid markers at the bottom of Fig. 2(b) and summarize the results in Table II. In both calculations, the obtained orbital multiplets deviated from our measurements, suggesting that the energy scales of local electronic correlations and bandwidth do not suffice to explain the insulating behavior of the CCDW phase. This inconsistency is in accordance with the experimental fact that the high symmetry phase of $1T$-TaS$_2$ is metallic, and that the insulating state only occurs in the presence of the periodic lattice distortion, in contrast to 3d transition metal compounds.

**Table II: Multiplet calculation results based on a single Ta atom with five d-orbitals.**

| Energy levels with SOC (eV) | d-d excitations with SOC (eV) | Energy levels without SOC (eV) | d-d excitations without SOC (eV) |
|---|---|---|---|
| 1.234 | 0.017/0.082 | 1.363 ($e_g$, 2-fold) | 0.106 |
| 1.316 | 0.289/0.371 | 1.469 ($a_{1g}$) | 2.774 |
| 1.605 | 2.65/2.667 | | |
| 4.255 | 2.939/2.956 | 4.243 ($e'_g$, 2-fold) | 2.88 |
| 4.272 | 3.021/3.038 | | |

To understand the role of the periodic lattice distortions in $1T$-TaS$_2$, we turn to a band-theoretical description of the excitations. We performed density functional theory (DFT) calculations for both the high-symmetry unit cell and the $\sqrt{13} \times \sqrt{13} \times 1$ supercell. We note the full CCDW has a supercell of $\sqrt{13} \times \sqrt{13} \times 3$. Our current selection of the supercell is sufficient to describe the in-plane electronic structure of $1T$-TaS$_2$, though the out-of-plane character remains metallic [28, 29, 30]. In order to properly describe the Ta d orbitals, we took the DFT+U [31] approach where we used the generalized gradient approximation (GGA) [32] for semi-local exchange correlation and a Hubbard U parameter for on-site electron interactions within Ta d orbitals. An effective Hubbard U of 2.27 eV was chosen, as determined previously from the first principles linear response theory [15, 29]. We discuss details of the DFT calculations in the **Supplementary Materials** [26]. The high-symmetry band dispersion is displayed in Fig. 3(a) and overlaid with the projection to the atomic orbital wavefunctions. It is apparent that the energy loss features observed in the experiment from 1.57 eV to 7.11 eV involve not only the Ta 5d-orbitals, but also possibly Ta 5p and the S 3p ligand states.

Using the linear response time dependent DFT [33], we compare the RIXS spectra with the energy loss function (ELF) calculated as the inverse of the dynamical dielectric function $\epsilon$ within the random phase approximation (RPA):

$$S(\mathbf{q}, \omega) = -Im(\epsilon^{-1}),$$

where

$$\epsilon(\vec{q}, \omega) = 1 - V_q \sum_k \frac{f_{k-q} - f_k}{\hbar(\omega + i\delta) + E_{k-q} - E_k}.$$

Details of the calculations are presented in the **Supplementary Materials** [26]. Notably, we did not take into account the effect of the core hole in our calculation as compared to other itinerant-electron based RIXS calculations [34, 35, 36], due to computational challenges associated with the size of the supercell. Following [34], we assume the core hole mainly acts as a scaling factor at resonance, which undoubtedly changes the relative spectral weight of the calculated peaks but is not expected to dramatically alter the position of these peaks. Therefore, we expect that our approach allows us to quantify the delocalized origin of the RIXS spectra and to disentangle the role of electronic states near the Fermi level from those of the core levels.

In Fig. 2(b), we show the calculated ELFs at momentum transfer $\mathbf{q} = (-0.1, 0, 0)$ for the high-symmetry and CCDW phases of 1$T$-TaS$_2$. The corresponding Brillouin zone (BZ) is defined from the high-symmetry unit cell for both phases. This corresponds to the total momentum transfer $\mathbf{Q} = (-1.1, 0, 6)$ where the RIXS data in Fig. 1 and Fig. 2(a) were taken. Here $\mathbf{Q} = \mathbf{q} + \mathbf{G}$ with $\mathbf{G}$ of reciprocal lattice vector. We chose this $\mathbf{Q}$ position as it is slightly away from the (-1, 0, 6) Bragg peak while keeping the RIXS spectrometer 90° relative to the incident X-ray to minimize the elastic peak intensity, following standard RIXS measurement protocols in the hard X-ray regime. Under the independent-particle approximation, a peak in the loss function is associated with a large number of electron-hole transition resonances (i.e., a large density of final states) sharing the same momentum transfer and energy difference. Calculated loss function away from the peaks comes from other electron-hole channels, albeit with smaller total transition probability than at the peak. Due to the itinerant nature of the electrons in this band-based approach, the calculated peaks on the energy loss function are generally broad, consistent with the observations in Table I. As shown in Fig. 2(b), both high-symmetry and CCDW phases have a prominent feature in their loss function between 2~3 eV. This increase in the loss function is associated with a kernel in the real part of the dielectric function (i.e., a plasmon) (Fig. S4 in the **Supplementary Materials** [26]). In contrast, features less than 1.5 eV appeared broader in the calculated loss function: these features are the result of an increase in the imaginary part associated with the interband transitions, with significant damping coming from the tail of the Drude peak. Hence our calculations show that both interband and plasmonic excitations impact the loss function in 1$T$-TaS$_2$, and that the interband excitations dominate the low-energy response and are strongly impacted by the structural distortion.

To clarify the detailed orbital nature of the energy loss features, we further decompose the contribution from relevant bands and atomic orbitals (Ta *5d* and S *3p*) to the calculated band structures (Fig. 3 (b)-(f)), by computing the transition intensities as the function of band index of all these excitations. This analysis, performed here for the high symmetry phase, allows us to assign orbital transitions to the observed RIXS features. The lowest energy loss feature around 1 eV is almost exclusively composed of the partially filled and the empty bands near the Fermi level, mainly within the partially-filled Ta $t_{2g}$ states ($d_{xy}$, $d_{yz}$ and $d_{zx}$, with *x, y, z* referring to the local coordinates within the TaS$_6$ octahedron [37, 38]), and is, logically, significantly impacted by the CCDW. The feature at 2 eV corresponds to the transitions from mixed S *3p* states to Ta $t_{2g}$ orbitals, which also make up the energy loss features at ~ 4 eV, ~ 5 eV. The $e_g$ states are only prominently involved in the ~ 7 eV feature. Hence, the manifold involved in the RIXS spectrum includes 3 Ta *5d* orbitals and 6 S *3p* orbitals, significantly departing from the pure *d-d* excitations observed in RIXS spectra in *3d* transition metal systems [19, 21, 22].

The assignment of orbital characters is consistent with the electron density of states measured from scanning tunneling spectroscopy (STS) [15]. Specifically, the STS resolved two sets of electronic states within $\pm 0.5$ eV of the Fermi level – localized electrons within $\pm 0.2$ eV that constitute the lower and upper Hubbard bands assuming a Mott insulator perspective, and the more extended electronics states $\pm 0.5$ eV and up. Both sets of states are centered around the Ta atom at the center of the CDW but show significant delocalization over the Ta clusters [29], resulting in smaller effective electronic interactions. Importantly, these electronic states contribute to the RIXS spectrum qualitatively different from the *3d* states in traditional Mott insulators and suggest that explanations of the RIXS spectra from the localized electron wavefunctions must involve the significant delocalization around the center of distortion.

Interestingly, the S ligand states are at least partially responsible for four of the five energy loss features, which was little discussed in previous literatures. The participation of ligand states is known to be vital in other quantum materials, e.g., cuprates, and most recently nickelates [23, 39]. Our result highlights the significance of ligand atoms in yet another quantum material and will affect the interpretations of the orbital textures in $1T$-TaS$_2$ and other dichalcogenides [15, 37, 38, 40].

### III. Discussion

We explore the temperature and momentum dependences of the orbital excitations in further details. In Fig. 4(a), we show the temperature dependence of the RIXS spectra at $Q$ = (-26/18 -1/9 6), which is a momentum transfer between M and K, across the CCDW-NCDW transition. This phase transition is of first order with over one order of magnitude change in resistivity. RIXS spectra were collected over both the warming and cooling cycles. We observed no appreciable changes through the phase transition. Note the charge orders in the CCDW and NCDW phases are very similar. Indeed, the NCDW phase is often interpreted as the CCDW domains separated by metallic states in some literatures [41]. In this sense, we do not expect a major change in the electron density of states and the orbital texture over the eV energy scale, except very near the Fermi level [30]. Thus, the persistence of the RIXS spectra can be naturally explained within the band-based framework.

Fig. 4(b) shows the RIXS spectra at base temperature along the high symmetry directions ($\Gamma - M - K - \Gamma - A$, defined in the inset of Fig. 4(c)) of reciprocal space, within the first 4 eV of energy loss. Using the same multiple peak fitting outlined previously, we analyzed the two lowest energy loss features and plot their dispersions in Fig. 4(c). Most prominent is a pronounced softening of the lowest energy loss feature from 1.3 eV to below 1 eV, over an extended region in the reciprocal space around $\Gamma$ and along $\Gamma - A$. In comparison, the second energy loss feature is generally flat over much of the BZ, apart from two $q$ points near but not at the M point.

The observed softening of the lowest energy loss feature is unexpected. In the majority of the existing RIXS literatures, orbital and charge transfer excitations over a few eV generally do not depend on the momentum transfer, especially in Mott insulators. Of the few notable exceptions, the Ruddlesden-Popper series of iridates Sr$_{n+1}$Ir$_n$O$_{3n+1}$ have dispersive orbital excitations between 0.5 eV and 1 eV from localized spin-orbital moments [5, 42]. In Sr$_2$CuO$_3$, the quasi-one-dimensional nature of the material gives rise to spin-charge separation and leads to several relatively well-defined bosonic modes between 1.5 eV and 3 eV [43]. In both examples, the electron wavefunctions are localized. For $1T$-TaS$_2$, due to the distortion of the TaS octahedron, there are 13 Ta $5d^1$ electrons per CCDW supercell in plane. Note the band folding along the c axis is not taken into account in the current calculations due to limited computing power. Given the electron delocalization, it is possible that the softening of the ~ 1eV feature arises from the further crystal field symmetry breaking and the splitting of the electron bands due to the stacking of the CCDW along the c axis [12, 40].

## IV. Conclusion

In sum, the orbital excitations in 1$T$-TaS$_2$ and the absence of excitations corresponding to the assumed Mott gap fundamentally disagree with the accepted scenario of localized electron wavefunctions. Instead, the orbital excitations can be described qualitatively and semi-quantitatively using extended electron wavefunctions. Our study reveals that the nature of the insulating state of 1$T$-TaS$_2$ significantly differs from canonical *3d* Mott insulators [19, 23], and underlies the importance of recent theoretical attempts to interpret energy loss features from a more itinerant picture in similar materials [33, 35, 44, 45, 46, 47, 48]. In these cases, the theoretical prediction required dedicated RIXS measurements either with high energy resolution challenging at today's instrumentation [44], or over large regions of the reciprocal space that cannot be reached especially for *3d* elements [45]. The case of 1$T$-TaS$_2$ shows unambiguously that the low-energy RIXS spectra should incorporate the distortion effect in addition to electronic correlations and will thus affect the understanding and interpretation of future experiments.

## Acknowledgements

The inception of the study, the analysis and interpretation of the experiment were supported by the U.S. Department of Energy, Office of Science, Basic Energy Sciences, Materials Science and Engineering Division. S.K acknowledges support from the Canada First Research Excellence Fund. The execution of the experiment was supported by Laboratory Directed Research and Development (LDRD) funding from Argonne National Laboratory, provided by the Director, Office of Science, of the U.S. Department of Energy under Contract No. DE-AC02-06CH11357. The density functional calculations were supported by the U.S. Department of Energy (DOE), Office of Science, Basic Energy Sciences Early Career Program under Award No. DE-SC0018080. This research used resources of the Advanced Photon Source, a U.S. Department of Energy (DOE) Office of Science User Facility operated for the DOE Office of Science by Argonne National Laboratory under Contract No. DE-AC02-06CH11357. Work performed at the Center for Nanoscale Materials, a U.S. Department of Energy Office of Science User Facility, was supported by the U.S. DOE, Office of Basic Energy Sciences, under Contract No. DE-AC02-06CH11357. Y.W. was supported by USTC Research Funds of the Double First-Class Initiative (No. YD9990002005). Y.L., W.L, X.L., and Y.S. thank the support from the National Key Research and Development Program under Contract No. 2021YFA1600201, the National Nature Science Foundation of China under Contract Nos. 11674326 and 11874357, the Joint Funds of the National Natural Science Foundation of China, and the Chinese Academy of Sciences' Large-Scale Scientific Facility under Contract Nos. U1832141, U1932217 and U2032215.

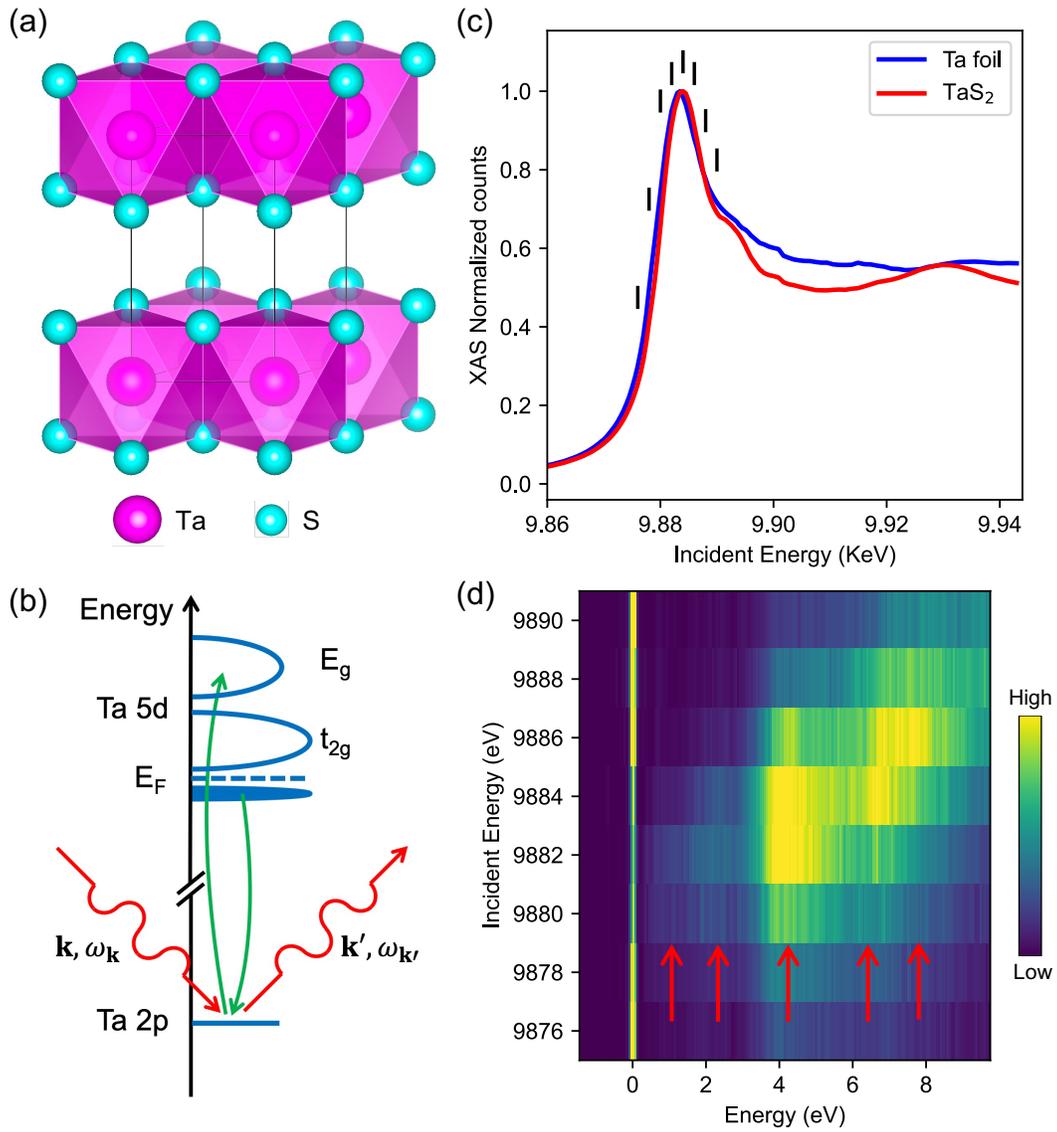

Figure 1. Energy loss spectra of 1$T$-TaS$_2$ across the Ta L$_3$ resonance. (a) Crystal structure of 1$T$-TaS$_2$. Each Ta atom is at the center of a Ta-S octahedron. (b) Schematics of the RIXS process at Ta L$_3$ edge. (c) X-ray absorption spectra at the Ta L$_3$ edge for 1T-TaS$_2$ and the reference Ta foil, respectively. The incident energies for the RIXS measurements in panel (d) are marked using black vertical lines. (d) RIXS spectra as a function of the incident energy across the Ta L$_3$ resonance at 10K. Red arrows are used as a guide to the eye to mark the energy loss features.

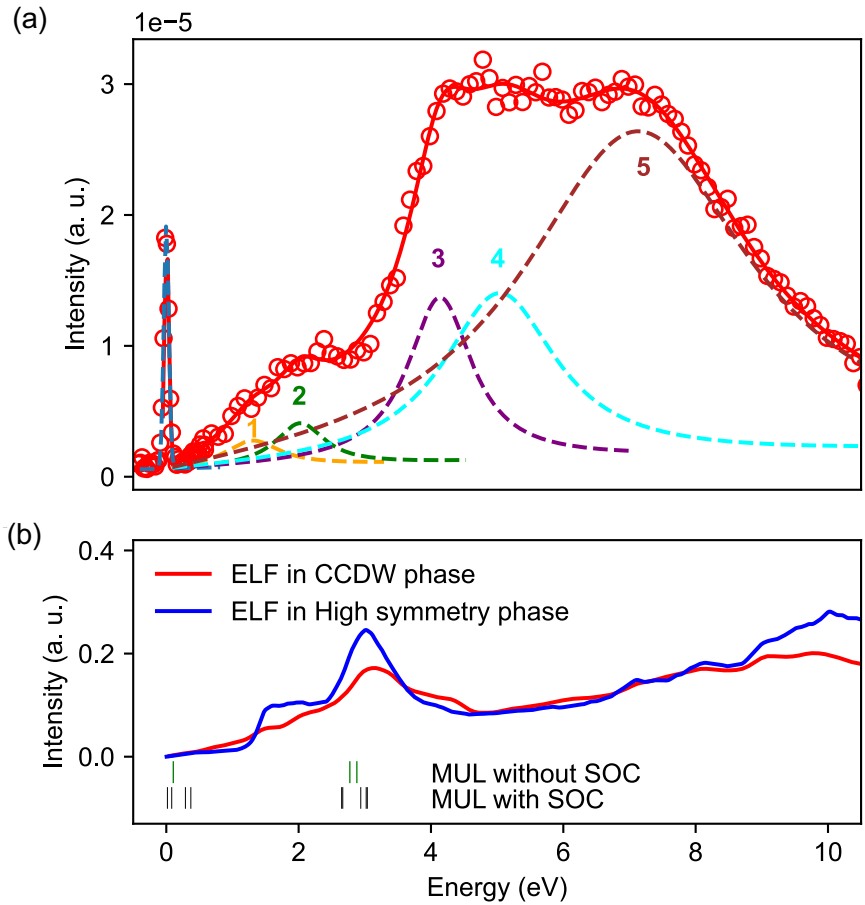

Figure 2. Comparing the measured and calculated energy loss features. (a) The RIXS spectrum at the incident energy of 9882 eV. The open red circles are the experimental data and the solid red line is the fit. The individual components of the fit corresponding to each energy loss feature are plotted using dashed lines. (b) The energy loss function (ELF) in the CCDW phase (red) and the high-symmetry phase (blue) calculated using DFT. The energies of the orbital excitations from the multiplet (MUL) calculations with and without SOC are marked using black and green vertical markers, respectively.

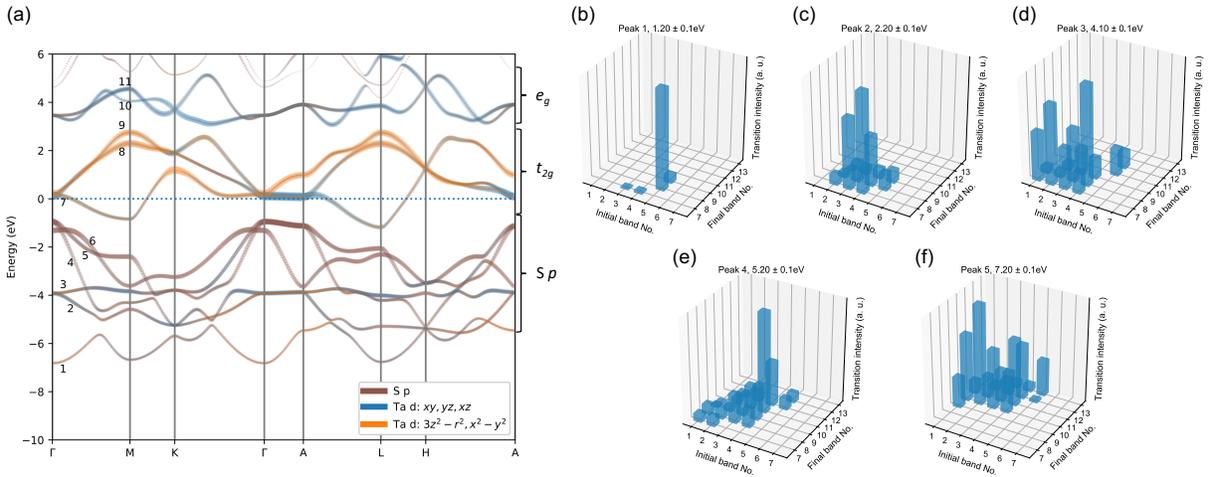

Figure 3. The high-symmetry band structure and the orbital-decomposed contribution to the RIXS energy loss features. (a) Band structure of the high-symmetry phase of $1T$-TaS$_2$ overlaid with projection to the atomic orbital wavefunctions. (b)-(f) Calculated orbital transition intensities as a function of initial and final band numbers at the energies of all five orbital excitations. The energy window is chosen to be $\pm 0.1$ eV for all the energy loss features.

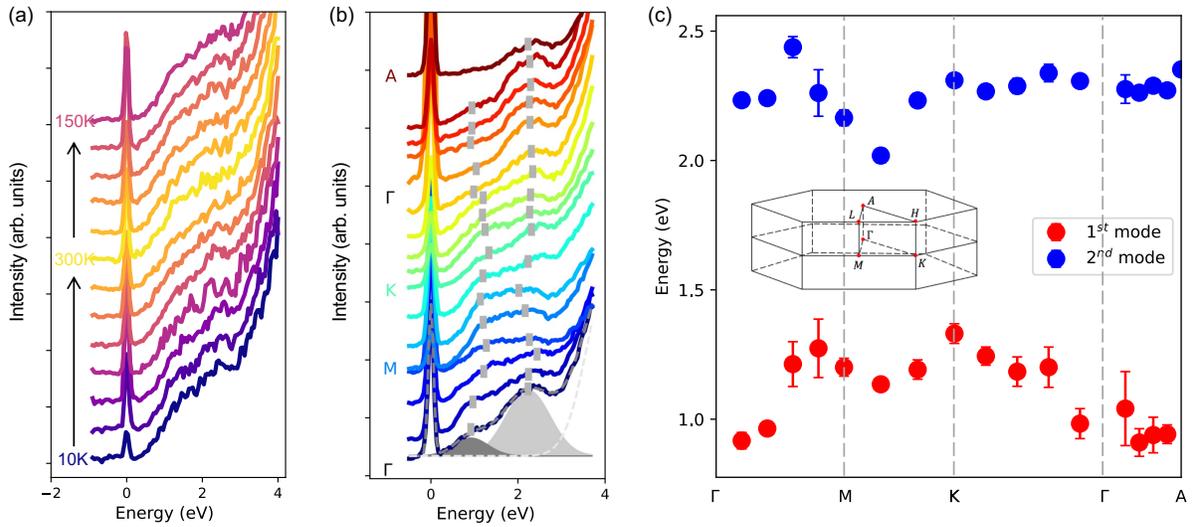

Figure 4. The temperature and momentum dependences of the lowest two orbital excitations. (a) The across the CCDW-NCDW phase transition, m dependence of the RIXS spectra within atures from the fit are displayed using grey th the contributions from the two orbital of the two lowest orbital excitations. Inset:

# Supplementary Materials for
# Interplay of broken symmetry and delocalized excitations in the insulating state of 1$T$-TaS$_2$

Xun Jia[1,*], Anubhab Haldar[2], Jungho Kim[3], Yilin Wang[4], Gilberto Fabbris[3], Karl Ludwig[5,6], Stefanos Kourtis[7], Mary Upton[3], Yu Liu[8], Wenjian Lu[8], Xuan Luo[8], Yu-Ping Sun[8,9,10], Diego Casa[3], Sahar Sharifzadeh[2,5,6], Pierre T. Darancet[11,*], Yue Cao[1,12,*]

[1] Materials Science Division, Argonne National Laboratory, Lemont, Illinois 60439, USA
[2] Department of Electrical and Computer Engineering, Boston University, Boston, MA, USA
[3] Advanced Photon Source, Argonne National Laboratory, Lemont, Illinois 60439, USA
[4] Hefei National Laboratory for Physical Science at Microscale, University of Science and Technology of China, Hefei 230026, China
[5] Department of Physics, Boston University, Boston, MA 02215, USA
[6] Division of Materials Science and Engineering, Boston University, Boston, MA, USA
[7] Institut quantique & Departement de physique, Universite de Sherbrooke, Quebec J1K 2R1, Canada
[8] Key Laboratory of Materials Physics, Institute of Solid State Physics, HFIPS, Chinese Academy of Sciences, Hefei 230031, China
[9] High Magnetic Laboratory, HFIPS, Chinese Academy of Sciences, Hefei 230031, China
[10] Collaborative Innovation Centre of Advanced Microstructures, Nanjing University, Nanjing 210093, China
[11] Center for Nanoscale Materials, Argonne National Laboratory, Lemont, Illinois 60439, USA
[12] Pritzker School of Molecular Engineering, University of Chicago, Chicago, Illinois 60637, USA

Emails: xjia@anl.gov (X.J.), ssharifz@bu.edu (S. S.), pdarancet@anl.gov (P.T.D.), yue.cao@anl.gov (Y.C.)

## S1. Sample synthesis

The high-quality single crystals of 1$T$-TaS$_2$ were grown utilizing the chemical vapor transport method with iodine as the transport agent. The high-purity Ta (3.5 N) and S (3.5 N) powders were mixed in chemical stoichiometry and heated at 850°C for about 100 hours in an evacuated quartz tube. The obtained TaS$_2$ powders and iodine (density: 5 mg/cm$^3$) were then sealed in another quartz tube and heated for 14 days in a two-zone furnace, in which the temperatures of both the source zone and the growth zone were settled at 900°C and 800°C, respectively. The tubes were rapidly quenched in cold water to guarantee retaining of the 1$T$ phase of TaS$_2$.

## S2. X-ray diffraction measurement

Single crystals of 1T-TaS$_2$ are aligned using a lab-based four-circle X-ray diffractometer with a photon energy of 8.05 keV at room temperature. The rocking curves of the (0 0 2) and (-1 0 4) Bragg peaks as well as a nearly commensurate charge order at (-0.245 -0.068 3.667) are displayed in **Figure S1** with the full-width-at-half-maximum (FWHM) of 0.256º, 0.245º and 0.244º, respectively. The FWHMs of all the peaks are limited by the mosaic of the crystal within the mm-size X-ray photon footprint.

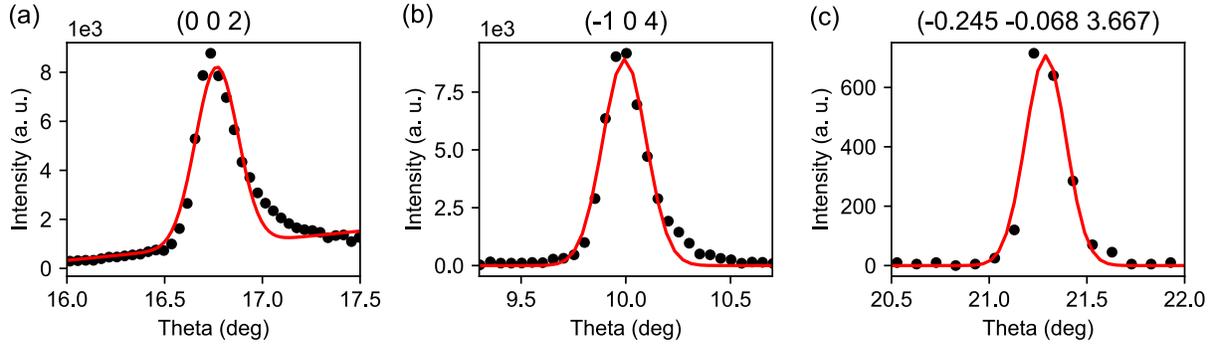

Figure S1. X-ray diffraction rocking curves of the single crystal 1T-TaS$_2$ at the (0 0 2), (-1 0 4) Bragg peaks and the charge order at (-0.245 -0.068 3.667). The data and the fits are shown using black dots and red solid lines respectively.

### S3. X-ray absorption spectroscopy (XAS) measurements

Ta L$_3$ XAS was measured at the 4-ID-D beamline of the Advanced Photon Source, Argonne National Laboratory. Pd mirrors were employed to both focus the X-rays to approximately $100 \times 200$ μm$^2$ and to reject higher harmonics. Data was collected at room temperature in partial fluorescence mode by monitoring the Ta Lα line intensity using a four elements silicon drift detector. The X-rays impinged on the sample at approximately 20° from the sample surface, with the fluorescence detector placed at 90° from the X-ray incidence.

### S4. RIXS measurements

The RIXS measurements were performed at the 27-ID-B (the MERIX endstation) of the Advanced Photon Source. The incident X-ray was monochromatized at the Ta L$_3$ edge. The scattered X-ray photons from the sample were reflected from a diced spherical Si (0, 6, 6) analyzer and collected using a one-dimensional MYTHEN detector with a pixel size of 50 μm. The sample, the analyzer crystal, and the detector were placed on a Rowland circle with a radius of 2 m in the vertical plane. The total energy resolution of the RIXS setup was ∼63 meV (see details in **S5. RIXS spectrometer setup**). The base temperature of the cryostat is between 10~15 K.

### S5. RIXS spectrometer setup

The total energy resolution of the RIXS setup comes from (1) the intrinsic Darwin width of the Si analyzer crystal $\Delta E_D = E_i \cot\Theta\, W_D$, (2) the geometrical energy resolution contributed from a finite photon footprint S$_1$ on the sample $\Delta E_{S1} = E_i \cot\Theta\, S_1/R$ and (3) the geometrical energy resolution contributed from a finite detector pixel size S$_2$ $\Delta E_{S2} = E_i \cot\Theta\, S_2/2R$, respectively [1, 2]. For Si (0 6 6), $W_D = 11.5$ μrad, $\Theta = 78.58°$, S$_1$ = S$_2$ = 50 μm and the arm radius R = 2 m. Thus $\Delta E_D$ = 22.9425 meV, $\Delta E_{S1}$ = 49.875 meV and $\Delta E_{S2}$= 24.9375 meV, respectively. The total energy resolution from the spectrometer is ∼ 60 meV. In **Figure S2**, we show the elastic line measured from a Ta-containing intermetallic compound. The total energy resolution (including from the incident X-ray and the spectrometer) is determined using FWHM from a Gaussian fit as ∼ 63 meV.

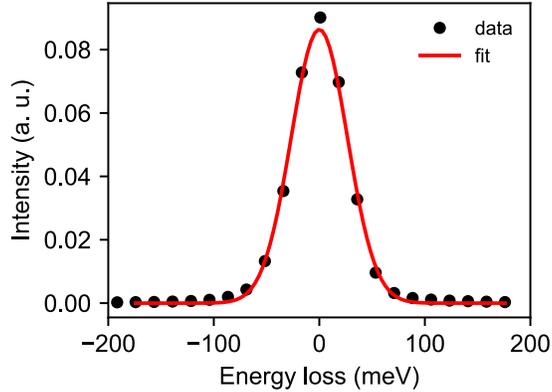

Figure S2. The elastic line measured at the Ta $L_3$ edge. The black dots and the red solid line are the data and the Gaussian fit, respectively.

### S6. High-resolution RIXS data at the Ta $L_2$ edge

We obtained high-resolution RIXS data at the Ta $L_2$ edge (**Figure S3**). The energy resolution is ~ 12 meV (**Figure S3 (a)**). Within the energy loss range of ~ 200 meV, and for the momentum transfer both at and away from the commensurate charge order (**Figure S3 (b)-(d)**), there is no clear signature of any lower-energy excitations e.g., phonons, or any features that could possibly arise from a Mott gap.

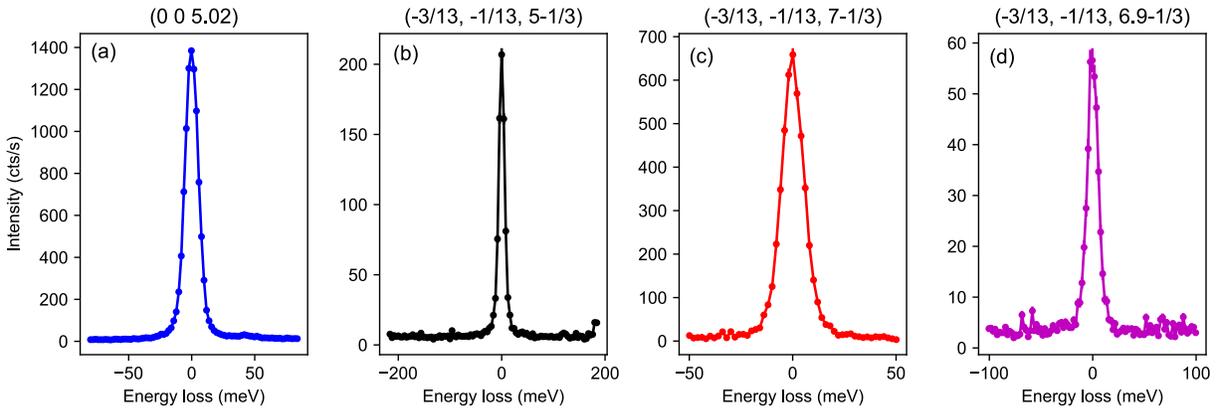

Figure S3. High-resolution RIXS data at the base temperature in the commensurate CDW phase obtained at the Ta $L_2$ edge.

### S7. Wannier tight-binding Hamiltonian for the multiplet calculations of 1$T$-TaS$_2$

We performed a density functional theory (DFT) combined with Wannier90 calculations to extract the strength of crystal field splitting (CFS) and spin-orbital coupling (SOC). The calculation was performed with the experimental crystal structure of 1$T$-TaS$_2$ in the primitive cell that contains one Ta site. The DFT part of the calculation with and without SOC was carried out using the Vienna Ab-initio Simulation Package (VASP) [3] with the projector augmented-wave (PAW) pseudopotential [4, 5] and. the Perdew-Burke-Ernzerhof (PBE) parametrization of the generalized gradient approximation (GGA-PBE) exchange-correlation functionals [6]. The energy cutoff of the plane-wave basis was set to be 500 eV, and a Γ-centered $11 \times 11 \times 11$ $\vec{k}$-point grid was used. A tight binding (TB) Hamiltonian consisting of all the Ta-5d orbitals was then obtained by the maximally localized Wannier functions method [7, 8] for the non-SOC case. The local on-site crystal field Hamiltonian took the following form (energy in eV, with respect to the Fermi level)

$$\hat{H}^{CF} = \begin{array}{c} d_{z^2} \\ d_{xz} \\ d_{yz} \\ d_{x^2-y^2} \\ d_{xy} \end{array} \begin{pmatrix} d_{z^2} & d_{xz} & d_{yz} & d_{x^2-y^2} & d_{xy} \\ 1.46949 & 0 & 0 & 0 & 0 \\ 0 & 3.373703 & 0 & 0 & -1.322224 \\ 0 & 0 & 3.373703 & -1.322224 & 0 \\ 0 & 0 & -1.322224 & 2.232118 & 0 \\ 0 & -1.322224 & 0 & 0 & 2.232118 \end{pmatrix}$$

We then diagonalized $\hat{H}^{CF}$ to get the crystal field levels listed in the first column, Table II.

To extract the strength of SOC, we added an atomic-type SOC Hamiltonian $\hat{H}^{SOC} = \vec{\lambda} \cdot \vec{l}$ to the TB model and determined λ by fitting the SOC band structures obtained from the DFT+SOC calculation. The result is λ ≈ 0.2 eV. We then diagonalized $\hat{H}^{CF} + \hat{H}^{SOC}$ to get all the 5d energy levels in the third column, Table II.

## S8. DFT-based calculations of the band structures and the RIXS spectra of 1$T$-TaS$_2$

All simulations were performed using the GPAW 19.8.1 [9, 10] package; the Atomic Simulation Environment (ASE) [11] 3.19.0 was used to set up the simulations. We performed first-principles density functional theory (DFT) and linear response time dependent DFT [12] calculations on a 1 × 1 × 1 unit cell and a $\sqrt{13} \times \sqrt{13} \times 1$ disordered supercell of TaS$_2$. In order to properly describe the Ta $5d$ orbitals, we took a DFT+U [13] approach where we used the GGA-PBE [6] for semi-local exchange correlation and a Hubbard U parameter for on-site electron interactions within the Ta $5d$ orbitals. We used an effective Hubbard U of 2.27 eV, as determined from first principles linear response theory previously [29]. PAW potentials [14, 15] were used to describe the core and nuclei of atoms as taken from the GPAW dataset version 0.9.2 with five and six electrons considered as valence for Ta and S, respectively.

The geometry of the unit cell was optimized with the exponential cell filter constraint [16] in ASE for the simultaneous optimization of the unit cell lattice constants and atomic positions, relaxing the *xx*, *yy*, and *xy* components of the strain tensor. The *z*-direction lattice constant was held fixed at an experimentally determined value of 5.8971 Å because there are no van der Waals interactions present in our DFT functional. The atomic positions were optimized until all forces were under 1 meV/Å. A Γ-centered $\vec{k}$-point grid of 14 × 14 × 8 and planewave energy cutoff of 950 eV were used, converging the total energy to < 1 meV/atom. The occupation smearing was set to a Methfessel-Paxton smearing [17] of order 1 and magnitude 0.1 eV.

For the supercell, the atomic positions and lattice vectors were iteratively optimized until forces were < 1 meV/Å. As is with the unit cell, the *xx*, *yy*, and *xy* components of the stress tensor were optimized. The *z*-axis was held fixed to the experimental value as is in the case of the unit cell. The $\vec{k}$-point mesh and planewave cutoff used for the supercell were 4 × 4 × 12 and 850 eV, respectively, converging the total energy to < 2 meV/atom. A Methfessel-Paxton smearing of order 0 and magnitude 0.05 eV was used. We used a lower smearing for the supercell to preserve the charge density wave state.

Density of states calculations were performed on a $\vec{k}$-point grid of 56 × 56 × 32 for the unit cell and 8 × 8 × 24 for the supercell with a reduced planewave cutoff of 350 eV to obtain the energies of states (bands) and Fermi-Dirac smearing of magnitude 0.1 eV. This energy cutoff converges the band energies to under 0.05 eV.

The energy loss function (ELF) of TaS$_2$ were calculated in the random phase approximation (RPA) [18, 19] as

where
$$S(\mathbf{q},\omega) = -Im(\epsilon^{-1}),$$

$$\epsilon^{-1} = \delta_{GG'} + \frac{4\pi}{|q+G|^2}\chi$$

is the inverse dielectric matrix, and the interacting density-density response function $\chi$ is related to the independent-particle density-density response by the Dyson-like relation,

$$\chi = \chi_0 + \chi_0 K \chi.$$

Here, $K = \frac{1}{|r-r|} + \frac{\partial V_{xc}[n]}{\partial n}$ is the interaction kernel consisting of the direct Coulomb and exchange-correlation terms. Within the RPA, we neglect the latter. The independent-particle density-density response (or susceptibility) is given by the expression,

$$\chi_0(q,\omega) = \sum_{k}^{BZ} \sum_{n,n'} \frac{f_{n,k} - f_{n',k+q}}{\omega + \epsilon_{n,k} - \epsilon_{n',k+q} + i\eta} \psi^*_{n,k}(r)\psi_{n',k+q}(r)\psi_{n,k}(r')\psi^*_{n',k+q}(r'),$$

where $\eta$ is a positive infinitesimal number and the f refer to the occupation numbers of the occupied (n) and unoccupied (n') state with energy $\epsilon_n$,k and $\epsilon_{n'}$,k+q at wavevectors **k** and **k** + **q**, respectively.

To assess the dynamical dielectric function, the number of unoccupied states used in the RPA were 32 and 196 for the unit cell and supercell, respectively. The GPAW package utilizes a multi-sampling scheme with a non-uniform frequency spacing. The spectrum was sampled on a grid where the energy spacing changed from an initial energy spacing of 0.05 eV to 0.1 eV continuously from 0 to 10 eV. The upper limit frequency for a converged Hilbert transformed spectrum for the real part of the dielectric function was set to 20 eV and local field effects were considered with plane waves up to 30 eV, corresponding to 294-297 planewaves depending on the point in the Brillouin Zone considered.

**S9. Real and Imaginary part of the calculated dielectric function.**

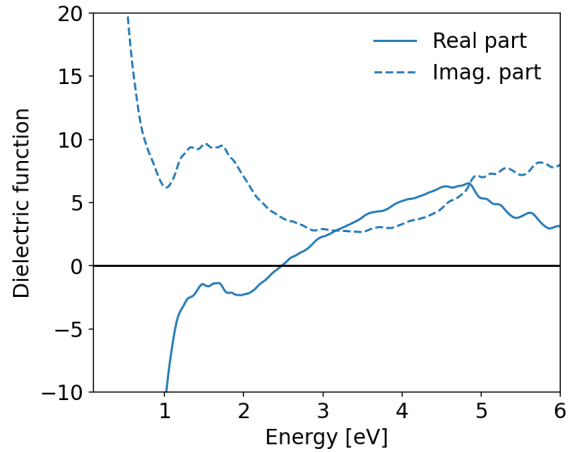

Figure S4. Real and imaginary part of the dielectric function in the CCDW phase of $1T$-TaS$_2$ at $q$ = (-0.1, 0, 0) calculated under the RPA. The solid and dashed lines correspond to the real and the imaginary part, respectively.

**Supplementary References:**